\documentclass[final]{jflart}
\usepackage[utf8]{inputenc}
\usepackage[T1]{fontenc}
\usepackage{color, colortbl}
\usepackage{graphicx}
\usepackage{mathrsfs}
\usepackage{float}
\usepackage{ulem}
\usepackage{csquotes}

\usepackage{./gospel}

\definecolor{thegray}{rgb}{0.9,0.9,0.9}
\definecolor{colorspec}{rgb}{0,0,0.797}
\definecolor{thered}{rgb}{0.797,0,0}
\definecolor{darkgreen}{rgb}{0.797,0,0}
\definecolor{theblue}{rgb}{0,0,0.797}
\definecolor{darkgray}{rgb}{0.8477,0.8477,0.8477}
\definecolor{ocaml-bg}{rgb}{0.9,0.9,0.9}
\definecolor{thegraygray}{rgb}{0.5,0.5,0.5}

\title{Le chameau et le serpent rentrent dans un bar : vérification
  quasi-automatique de code OCaml en logique de séparation}

\titlerunning{}

\author[1,2]{Charlène Gros}
\author[2]{Mário Pereira}
\authorrunning{Gros, Pereira}

\affil[1]{Junia ISEN Lille, Lille, 59000, France — Tarides, Paris, 75005, France}
\affil[2]{Nova School of Science and Technology, NOVA LINCS, Portugal}

\begin{document}

\maketitle

\begin{abstract}
  Cet article présente une traduction de programmes OCaml spécifiés en
  Gospel vers Viper, un langage intermédiaire de preuves formelles
  supportant la logique de séparation.
  L'objectif pratique est d'ajouter à Cameleer un nouveau
  \textit{backend} pour prouver des
  programmes OCaml manipulant le tas.
  La spécification logique de tels
  programmes OCaml est décrite dans le langage Gospel
  et nous détaillons ici les extensions apportées
  pour le support de la logique de séparation en Viper.
\end{abstract}

\section{Introduction}

Gospel (Generic OCaml Specification Language)~\cite{GOSPEL} est un
langage de spécification conçu pour OCaml permettant l'annotation de
programmes avec des contrats formels.
Celles-ci spécifient les
invariants des types, les pré-conditions et post-conditions des
fonctions, les effets, ainsi que les exceptions levées par une
fonction. La syntaxe non intrusive de Gospel facilite la génération de
documentation détaillée, permet la génération de tests à l’aide de
l'outil Ortac~\cite{DBLP:conf/rv/FilliatreP21} et intègre un support
de preuve formelle \emph{via} Cameleer~\cite{DBLP:conf/cav/PereiraR20}
pour les programmes OCaml.

Les annotations Gospel offrent un moyen fiable de formaliser les
comportements attendus d'un programme OCaml. Elles sont intégrées dans
le code source, ce qui fournit une description claire et non ambiguë du
comportement des modules et des fonctions, garantissant ainsi que la
documentation reste en phase avec l'implémentation.
Cela facilite la compréhension et la maintenance du
code en offrant une référence formelle aux développeurs.

Pour la vérification formelle des programmes OCaml, l'outil Cameleer
s'appuie sur Gospel pour annoter le code OCaml et le traduit ensuite
dans le langage WhyML, utilisé par la plateforme de vérification
déductive Why3~\cite{DBLP:conf/esop/FilliatreP13}.
Cette dernière s'appuie sur
le calcul des plus faibles pré-conditions, une méthode de vérification
déductive qui génère des conjectures logiques pour chaque programme et qui,
une fois vérifiées, garantissent qu'un programme est
conforme à sa spécification. Cette méthode permet de démontrer des
propriétés comme les invariants et les conditions de validité d'un
programme, en produisant des obligations de preuve transmises à des
solveurs automatiques tels qu'Alt-Ergo, Z3, CVC5 et bien d'autres.
Toutefois, bien que Why3 gère une forme de mutabilité, elle est limitée
lorsqu'il s'agit de traiter des structures mutables récursives. En
effet, le calcul des plus faibles pré-conditions et le système de
types et d'effets de Why3~\cite{filliatre:hal-01256434} imposent la
gestion statique de l'\emph{aliasing}, ce qui empêche notamment
l'utilisation de champs à la fois récursifs et mutables au sein de
types ou de structures de données.
L'exemple présenté ci-dessous illustre la définition
d'une liste chaînée en WhyML.
\begin{why}
type t = Nil | Cons cell
with cell = { mutable content: int; mutable next: cell }
  \end{why}
\vspace{-3ex}
\begin{lstlisting}[escapeinside={*?}{?*},basicstyle=\ttfamily,%
  columns=flexible,backgroundcolor=\color{ocaml-bg}]
  *?{\color{thered}{Error:}}?* Recursive types cannot have mutable fields
\end{lstlisting}
Le type \texttt{cell} est rejeté par Why3, qui interdit
un type récursif à contenir un champ mutable pointant vers lui-même.

Les outils basés sur la \emph{logique de séparation}~\cite{10.5555/645683.664578}
permettent de résoudre une
grande partie de ces problèmes. Le principe fondamental est de
raisonner sur des portions disjointes de mémoire à l'aide de
l’opérateur de conjonction séparante.
Bien que Why3 impose des restrictions via un suivi statique
des alias, ce qui permet de déterminer précisément quelles
modifications affectent une valeur mutable,
il ne peut pas exprimer explicitement la séparation ou la propriété
(\emph{ownership}) de ces portions de mémoire.
En revanche, la logique de séparation garantit grâce à la
distinction des portions de mémoire et la fameuse \emph{frame rule},
qu'aucune modification externe n'a eu lieu. L'intégration de cette
logique comme nouveau \emph{backend} dans Cameleer permet d'élargir la
vérification à un éventail plus vaste de programmes OCaml annotés avec
Gospel.

Un mécanisme de traduction de code OCaml annoté avec Gospel vers la
logique de séparation est déjà en cours de développement\footnote{voir
  \url{https://github.com/ocaml-gospel/gospel2cfml}}. Celui-ci traduit
le programme OCaml annoté en CFML~\cite{10.1145/2034773.2034828}, une
extension de Rocq (anciennement Coq) dédiée à la logique de
séparation. Cependant, faire une preuve avec CFML demande toujours une
connaissance de l'assistant de preuve Rocq et cette
traduction est limitée aux interfaces OCaml.


L'extension proposée ici est une traduction vers le langage
intermédiaire Viper, qui nous permet de conduire une preuve
quasi-automatique de la correction d'un programme. Viper (Verification
Infrastructure for Permission-based
Reasoning)~\cite{DBLP:conf/vmcai/0001SS16} est un outil basé sur la
logique de séparation et une de ses variantes : les \emph{Implicit
  Dynamic Frames}~\cite{Smans2009ImplicitDF, ParkinsonSummers12}.  Il
propose un langage intermédiaire ainsi que deux \emph{backends} de
vérification : l'un basé sur l'exécution symbolique et l'autre sur la
génération de conditions de vérification. Ces deux approches
s'appuient sur le solveur SMT \textquote{Z3} pour résoudre les
obligations de preuve. Le langage intermédiaire Viper est un langage
impératif et séquentiel, offrant des fonctionnalités de haut niveau
pour formuler des problèmes de vérification, tout en étant
suffisamment flexible pour automatiser la vérification dans divers
langages de programmation. Il est notamment utilisé pour Rust,
Python et Go à travers les \emph{frontends} respectifs
Prusti~\cite{DBLP:journals/pacmpl/Astrauskas0PS19},
Nagini~\cite{DBLP:conf/cav/Eilers018}, et
Gobra~\cite{DBLP:conf/cav/WolfACOPM21}.

Le développement présenté dans cet article est une extension de
l'outil Cameleer pour traduire les programmes OCaml annotés vers
Viper. Dans la suite nous nous référerons à cette extension sous le
nom de \texttt{gospel2viper}.

\paragraph{Remarque préliminaire.} L'extension présentée ici est un
travail toujours en cours de développement. Nous ne traitons qu'un
sous-ensemble réduit d'OCaml, qu'on pourrait dire proche
d'\texttt{ocaml-light} : en particulier, définitions et appels de
fonctions, variables locales, type algébriques (mais avec des \emph{records} en
argument), structure conditionnelle, filtrage, affectation de champs
mutables. Par conséquent, nous ne supportons ni les boucles ni les
tableaux. En ce qui concerne les définitions de types, nous ne
traitons pas les types polymorphiques. Nous n'autorisons, pour
l'instant, que des entiers comme type des éléments d'une structure de
données. Enfin, côté spécification, nous n'utilisons que des
\emph{séquences mathématiques} comme modèle logique.


\section{Présentation de \texttt{gospel2viper}}


Actuellement, Cameleer prend en charge la traduction vers WhyML,
utilisé par Why3. L'extension \texttt{gospel2viper} vise à résoudre
certaines limitations imposées par l'utilisation de Why3. Il est
toutefois nécessaire d'étendre aussi le langage Gospel avec des
constructions classiques de logique séparation, \emph{e.g.}, la notion
de possession des champs d'un \emph{record}. 
%
%
Dans cette section nous introduisons au fur et à mesure les concepts
les plus importants de Gospel et Viper utilisés dans notre
traduction. L'explication est complétée dans la
Section~\ref{sec:etude-de-cas}, lorsque ces mêmes concepts sont
utilisés dans une étude de cas.

%

En Viper, les données mutables sont des références déclarées en tant
que champs par le mot-clé \texttt{field}, et les structures de données sont décrites par des
assertions dans des prédicats de représentation \emph{via} le mot-clé
\texttt{predicate}. C'est dans ce prédicat qu'est notamment décrit le
\textit{footprint} de mémoire d'une structure, ainsi que sa relation
avec un modèle logique.
L'opérateur \texttt{\&\&} agit comme la conjonction séparante de la
logique de séparation. En tant qu'abstraction de son contenu, Viper
impose l'ouverture explicite d'un prédicat de représentation
par l'opération de dépliage (\texttt{unfold}), afin de
révéler son contenu et accéder aux permissions en mémoire.
La fonction opposée est celle de pliage (\texttt{fold}), utilisée
pour abstraire le prédicat de nouveau.




La traduction de certains éléments d'OCaml vers Viper se révèle
triviale, comme la distinction effectuée par Viper entre les
\texttt{methods} et les \texttt{functions}, qui correspond
respectivement aux fonctions OCaml et aux fonctions logiques Gospel
définies par le mot-clé \texttt{function}. Le \emph{backend} proposé
ici utilise l'AST (\emph{Abstract Syntax Tree}) de Cameleer, lui-même basé sur l'AST non-typé
d'OCaml. Bien que le langage Viper nécessite les types explicites,
nous demandons pour le moment de les inclure dans le code OCaml et
Gospel. L'ajout de l'inférence de type au sein de l'extension viendra
dans le futur.

Les paragraphes suivants détaillent les spécificités du langage Viper,
ainsi que les moyens mis en place pour trouver l'ensemble des
informations dans le code Cameleer pour procéder à la traduction.
Tous les exemples affichent à gauche le code Cameleer et à droite
leur traduction en Viper.

\paragraph{Références en Viper.} Pour créer une ressource accessible
via une référence en Viper, celle-ci doit être déclarée comme un
\texttt{field}. Ensuite, chaque variable de type \texttt{Ref} pourra
accéder à l'entièreté des champs déclarés, car il n'y a aucune notion
d'encapsulation en Viper.
Cependant, pour préciser quels champs sont utilisés par une ressource donnée,
il est nécessaire de demander explicitement les permissions correspondantes
à l'aide de l'instance \texttt{acc{}}, comme illustré par la suite.
La définition de champs est
naturellement décrite en OCaml par un \emph{record}, ce qui nous permet
d'avoir toutes les informations sur les types des différents \emph{fields}
Viper.


\paragraph{Encodage des types algébriques en Viper.}
Prenons le
type OCaml suivant. Il reprèsente une liste chainée avec deux champs mutables.
\begin{gospel}
type cell = Nil | Cons { mutable content: int; mutable next: cell }
\end{gospel}
Comme Viper fournit la constante \texttt{null}
pour le type \texttt{Ref}, il serait tentant de traduire le constructeur
\texttt{Nil} d'OCaml vers \texttt{null}. Néanmoins, une telle approche
ne serait pas satisfaisante quand une définition de type OCaml
présente plusieurs cas de base.
Une extension à Viper a été récemment proposée pour encoder dans ce
langage une notion de type
algébrique~\cite{maissen2022adding}. C'est exactement cette approche
que nous utilisons dans \texttt{gospel2viper}. Cela permet de décrire de
manière très similaire en Viper la structure d'un type OCaml, comme
montré en Figure~\ref{fig:adt-list} pour la liste chaînée.


\begin{figure}
\begin{minipage}[t]{0.5\textwidth}
  \centering
  \begin{gospel}
type cell =
  | Nil
  | Cons of { mutable content : int;
              mutable next : cell }
  \end{gospel}
  \end{minipage}
  \hfill
  \begin{minipage}[t]{0.5\textwidth}
      \centering
      \begin{viperlang}
adt Cell {
  Nil()
  Cons(cell: Ref)
}
field content : Int
field next: Cell
\end{viperlang}
\end{minipage}
\caption{Encodage du type de listes chaînées à l'aide des ADTs en Viper.}
\label{fig:adt-list}
\end{figure}

En Viper, nous créons une instance d'un ADT (\emph{Algebraic DataTypes}) \texttt{Cell}, qui est
soit \texttt{Nil()} (pas d'argument), soit \texttt{Cons} avec un
argument de type \texttt{Ref} nommé ici \texttt{cell} (tout autre nom
serait également acceptable). Enfin, nous pouvons introduire les deux
champs \texttt{content} et \texttt{next}, que nous utiliserons uniquement
lorsqu'une valeur du type \texttt{Cell} sera non-\texttt{Nil}. Il est
aussi important de mentionner que cette utilisation de la construction
\texttt{adt} en Viper nous donne automatiquement deux destructeurs
mutuellement exclusifs d'une \texttt{Cell} : \texttt{isNil} et
\texttt{isCons}. Par exemple, pour un certain~\texttt{c} de type
\texttt{Cell}, si \texttt{c.isNil} se vérifie, alors Viper nous empêche
d'accéder au champ \texttt{next} de~\texttt{c}.
Néanmoins, le filtrage par motif
(\emph{pattern-matching}) doit encore être géré manuellement en utilisant ces
mêmes destructeurs dans une séquence imbriquée de \texttt{if..else}.

\paragraph{La possession d'un champ mutable.} En Viper, le
prédicat d'accessibilité \texttt{acc} est utilisé pour la gestion
des permissions d'accès aux champs d'un objet. Chaque opération, qu'il
s'agisse d'une expression, d'une assertion ou d'une modification
d'état, nécessite explicitement une permission afin d'accéder aux champs
du programme. Ceci permet de déterminer quelles portions de la
mémoire sont affectées par une opération. Les permissions garantissent
la sécurité et l'intégrité du programme, notamment au travers du
mécanisme de \emph{framing}, qui permet de préserver des propriétés
sur la mémoire entre des appels de méthodes ou des opérations
distinctes. L'utilisation de \texttt{acc(x.f)} dans une pré-condition
signifie que l'appelant d'une méthode doit céder temporairement le
droit d'accès au champ \texttt{x.f} à la méthode appelée. En
contrepartie, l'apparition de \texttt{acc(x.f)} dans une
post-condition indique que cette permission sera rendue à l'appelant
une fois l'opération terminée.  Si plusieurs champs demandent la
permission, les prédicats d'accéssibilité \texttt{acc} seront séparés
par l'opérateur \texttt{\&\&}, qui sera compris en tant que
conjonction séparante. Cela permet de garantir que les opérations
n'interfèrent pas entre elles et ne modifient pas des champs pour
lesquels elles n'ont pas explicitement obtenu l'accès, assurant ainsi
une gestion fine des ressources et une absence d'\emph{aliasing} entre
les différentes parties du programme. Pour couvrir cette nouvelle idée
qui n'existe pas encore dans Gospel, nous avons ajouté l'opérateur
\emph{wiggle arrow} ($\leadsto$).
Ci-dessous, l'annotation Gospel (à gauche) et sa traduction Viper (à droite) :

 \begin{minipage}[t]{0.5\textwidth}
  \centering
  \begin{gospel}
c *?$\leadsto$?* {content; next}
  \end{gospel}
  \end{minipage}
  \hfill
  \begin{minipage}[t]{0.5\textwidth}
      \centering
      \begin{viperlang}
acc(c.content) && acc(c.next)
\end{viperlang}
\end{minipage}
Cet opérateur permet de spécifier de manière précise la possession d'un \emph{record} et
la possession individuelle de chaque champ. Cette notion de possession
des champs de \emph{record}, ainsi que l'opérateur ($\leadsto$), sont
fortement inspirées par CFML.

\paragraph{Définition de prédicats de représentation.} En Viper, un
prédicat de représentation est un mécanisme d'abstraction sur les
assertions, y compris sur les ressources telles que les permissions
d'accès à des champs. Il permet de regrouper sous une forme compacte
et paramétrée des propriétés et permissions liées à des structures de
données potentiellement récursives. Contrairement aux fonctions, les
prédicats ne sont pas automatiquement développés : leur corps qui
constitue une assertion doit être explicitement révélé à l’aide de
l'instruction \texttt{unfold}, ou masqué avec l'instruction
\texttt{fold}.
Ces prédicats sont définis en Gospel par le mot-clé \texttt{predicate}
en exprimant les mêmes assertions.

\paragraph{Opérations \texttt{fold} / \texttt{unfold} et
  \texttt{apply}.} 
%
L'opération \texttt{unfold} modifie l'état du programme en remplaçant
la ressource du prédicat par les assertions de son corps, donnant
alors l'accès aux \emph{fields}. Une variante de cette opération
existe pour temporairement exposer l'intérieur du prédicat, avec
\texttt{unfolding ... in}.  L'opération inverse, \texttt{fold},
remplace les assertions par une instance du prédicat.
Cette dernière, notée \texttt{P($e_1$, \dots, $e_n$)}, est une assertion de
ressource dans Viper qui peut être consommée ou produite.

Ci-dessous, nous présentons l'implémentation d'une
méthode Viper simple pour changer la valeur d'une référence,
en omissant les permissions sur les champs.
  \begin{viperlang}
field v: Int
predicate P (c: Ref) { // pr*?é?*dicat cachant la possession du champ v }
method foo (c: Ref)
  requires P(c)
  ensures  P(c)
    { *?\uwave{\texttt{\phantom{l}c.v := 0\phantom{l}}}?* }
  \end{viperlang}
\vspace{-3ex}
\begin{lstlisting}[escapeinside={*?}{?*},basicstyle=\ttfamily,%
  columns=flexible,backgroundcolor=\color{ocaml-bg}]
*?{\color{thered}{Error:}}?* Assignment might fail. There might be insufficient permission to
        access c.v
\end{lstlisting}
Viper
indique que cela peut échouer car il n'a pas la permission d'accès au
champ \texttt{c.v}. Pour que Viper puisse prouver cette
implémentation, il faut ajouter en première instruction le dépliage de
\texttt{P} pour y exposer les permissions. À la fin, il faut utiliser
\texttt{fold} pour rétablir le prédicat. Sans ces appels à
\texttt{fold} et \texttt{unfold}, Viper n'est pas capable de suivre
l'évolution de qui possède les permissions, et donc ne peut assurer
que les modifications n'ont pas changé un autre morceau de la mémoire.

À ce jour, aucune solution de génération automatique de ces appels
n'est connue, ce qui oblige les programmeurs Viper à les spécifier
manuellement. Par conséquent, il est également nécessaire d'inclure
dans les annotations Gospel cette notion, car l'utilisateur n'est
pas censé toucher à la traduction Viper. 

Pour permettre l'utilisation de lemmes auxiliaires traduits sous
forme de méthodes en Viper, nous avons introduit l'opération
\texttt{apply} dans Gospel. Elle permet d'appeler ces lemmes directement
depuis les annotations sans interférer avec le code OCaml, bien qu'elles
doivent être placées dans l'implémentation OCaml, tout comme les appels à
\texttt{fold} et \texttt{unfold}. 

\section{Étude de cas : module \texttt{Queue} d'OCaml}
\label{sec:etude-de-cas}

Dans cette section, nous présentons la vérification formelle du module
\texttt{Queue} de la bibliothèque standard d'OCaml. Dans cette implémentation, le
type des files utilise des références pour avoir des opérations
d'ajout, de concaténation et de retrait en temps constant. 
Nous présentons à gauche le code OCaml annoté et à droite sa
traduction en Viper générée par notre extension.

\paragraph{Définition de segments de liste.} En utilisant l'encodage
des listes chaînées comme montré dans la Figure~\ref{fig:adt-list},
nous introduisons le prédicat \texttt{cellSeg} comme suit :

\begin{minipage}[t]{0.5\textwidth}
  \centering
\begin{gospel}
(*@ predicate cellSeg
    (from: cell)
    (v: int sequence)
    (to: cell)
   = if v = empty then to = from
     else
       let Cons c = from in
       c *?$\leadsto$?* {content; next} &&
       c.content = v[0] &&
       cellSeg c.next (v[1 ..]) to *)
\end{gospel}
  \end{minipage}
  \hfill
  \begin{minipage}[t]{0.5\textwidth}
      \centering
\begin{viperlang}
predicate CellSeg (from: Cell,
  v: Seq[Int], to: Cell)
{ v == Seq[Int]() ?
    to == from
  : from.isCons &&
    let c == (from.cell) in
    acc(c.content) &&
    acc(c.next) &&
    c.content == v[0] &&
    CellSeg(c.next, v[1 ..], to) }
\end{viperlang}
\end{minipage}
%
%
%
%
Ce prédicat \texttt{cellSeg} (pour \textquote{segment de cellules})
décrit une succession de cellules entre deux références :
\texttt{from} et \texttt{to}, ayant pour modèle la séquence d'entiers
\texttt{v}. Ce prédicat vérifie dans un premier temps que si le
modèle est vide alors il n'y a aucune cellule entre les deux bornes
\texttt{from} et \texttt{to}. Sinon, nous avons accès aux champs
\texttt{content} et \texttt{next} de l'élément courant et nous
continuons la récursion sur la cellule suivante jusqu'à atteindre la
fin du segment. À chaque itération, il est vérifié que le contenu
courant de la cellule est bien le même que la tête du modèle, lui-même
dépilé lors de l'appel récursif. L'accès aux champs \texttt{content}
et \texttt{next} par la cellule \texttt{from} est modélisé par le
nouvel opérateur ($\leadsto$) en Gospel. Il se traduit en un appel au
prédicat d'accessibilité \texttt{acc} de Viper et permet de demander
le transfert des permissions afin de pouvoir lire les valeurs des
champs demandés.

Remarquons la construction \texttt{let Cons c = from in} : sa
sémantique est que \texttt{from} est forcément une cellule de la forme
\texttt{Cons}. Cela se traduit en Viper par le destructeur
\texttt{from.isCons}.

\paragraph{Définition de la structure de file.}
Nous présentons en
ci-dessous le type \texttt{queue} et le prédicat de
représentation de la file.

\begin{minipage}[t]{0.5\textwidth}
  \centering
\begin{gospel}
type queue = {
  mutable length : int;
  mutable first : cell;
  mutable last  : cell
}

(*@ predicate queue (q: queue)
    (v: int sequence) =
  q *?$\leadsto$?* {length; first; last} &&
  if v = empty then
    q.first = Nil && q.last = Nil &&
    q.length = 0
  else
    length v = q.length &&
    cellSeg q.first (drop_last v)
            q.last &&
    cellSeg q.last (take_last v) Nil *)
\end{gospel}
  \end{minipage}
  \hfill
  \begin{minipage}[t]{0.5\textwidth}
      \centering
\begin{viperlang}
field length: Int
field first: Cell
field last:  Cell

predicate Queue (q: Ref, v: Seq[Int])
{
  acc(q.length) && acc(q.first) &&
  acc(q.last) &&
  (v == Seq[Int]()
   ? q.first.isNil && q.last.isNil
     q.length == 0
   : |v| == q.length &&
     CellSeg(q.first, drop_last(v),
             q.last) &&
     CellSeg(q.last, take_last(v),
             Nil()))
}
\end{viperlang}
\end{minipage}
Elle est représentée par trois champs
mutables : \texttt{length} qui stocke la longueur de la file,
\texttt{first} qui est la première cellule de la file et
\texttt{last} qui est sa dernière cellule. Le prédicat \texttt{queue}
décrit l'état de la file par rapport au modèle \texttt{v}. Ensuite,
deux cas sont possibles une fois les accès en mémoire demandés : si la
file est vide alors les cellules de début et de fin sont \texttt{Nil},
ainsi que la taille vaut zéro ; sinon, les tailles doivent être les
mêmes, et l'on asserte que le segment de cellules de la première à
l'avant dernière est bien formé et que la dernière cellule contient
effectivement le dernier élément du modèle.

\paragraph{Création de file.} La fonction \texttt{create} est utilisée pour créer
une nouvelle file vide. Voici son implémentation et sa spécification
logique :

\begin{minipage}[t]{0.5\textwidth}
  \centering
\begin{gospel}
let create () : queue =
  let q : queue =
    Cons { length = 0;
           first  = Nil;
           last   = Nil } in
  (*@ fold queue q empty *)
  q
(*@ q = create ()
      ensures queue q empty *)
\end{gospel}
  \end{minipage}
  \hfill
  \begin{minipage}[t]{0.5\textwidth}
      \centering
\begin{viperlang}
method create () returns (q: Ref)
  ensures Queue(q, Seq[Int]())
{
  q := new(length, first, last)
  q.length := 0
  q.first := Nil();
  q.last := Nil();
  fold Queue(q, Seq[Int]())
}
\end{viperlang}
\end{minipage}

La post-condition de cette fonction spécifie
qu'à la fin de son exécution, la file renvoyée sera bien formée.
C'est l'instruction \texttt{let ... = ... in} qui remplit le
\emph{record}. Elle est scindée en deux en Viper: en premier, l'instruction
\texttt{new\{...\}} va assigner une nouvelle référence pour
\texttt{q} et aspirer les permissions d'écriture des champs entre
accolades. Ces permissions sont ensuite expirées \emph{via}
l'instruction \texttt{fold}, qui atteste que la file fraîchement crée
est vide et bien formée. Ensuite les champs sont initialisés.

\paragraph{Ajout d'un élément.} Nous continuons cette étude de cas par
la présentation de la spécification de la fonction qui ajoute un
élément à la fin de la file. L'intégralité de cette preuve ainsi que
sa traduction en Viper est entièrement disponible en
ligne\footnote{\url{https://github.com/ocaml-gospel/gospel2viper/blob/main/example}}.

Pour raisonner sur la concaténation de plusieurs segments de listes,
nous devons faire appel à des lemmes classiques de logique de
séparation~\cite{10.1145/2854065.2854068}. En particulier, le lemme
sur la transitivité de la concaténation de trois segments de liste
que l'on retrouve en Gospel de la manière suivante :

\begin{viperlang}
(*@ lemma cellSeg_trans (c1: cell) (c2: cell) (c3: cell)
             (v1: int sequence) (v2: int sequence) (v3: int sequence)
    requires cellSeg(c1, v1, c2) && cellSeg(c2, v2, c3) &&
             cellSeg(c3, v3, Nil)
    ensures  cellSeg(c1, v1 ++ v2, c3) && cellSeg(c3, v3, Nil) *)
\end{viperlang}
Avec ce lemme nous allons pouvoir montrer qu'à chaque
fois que l'on ajoute un élément à la fin de la file, cela va créer un
nouveau segment de liste qui sera ensuite concaténé au segment de
liste partant de la référence \texttt{last}.
La spécification de la fonction d'ajout est simple et sa traduction
est triviale, comme visible ci-dessous :

\begin{minipage}[t]{0.5\textwidth}
  \centering
\begin{gospel}
let add (q: queue) (x: int) = ...
(*@ add x q [v: int sequence]
    requires queue q v
    ensures  queue q (v ++
             (singleton x)) *)
\end{gospel}
\end{minipage}
\hfill
\begin{minipage}[t]{0.48\textwidth}
            \centering
\begin{viperlang}
method add (q: Ref, x: Int, v: Seq[Int])
  requires Queue(q, v)
  ensures  Queue(q, v ++ Seq(x))
{ ... }
\end{viperlang}
\end{minipage}
En entrée la fonction reçoit une file valide et doit en renvoyer une
dans laquelle
la valeur \texttt{x} aura été ajoutée.
La
fonction logique \texttt{singleton} renvoie une séquence qui ne
contient que l'élément passé comme argument. Nous utilisons les
arguments \emph{ghost} de Gospel pour transmettre le modèle à la
fonction. 
Cet argument \emph{ghost}, ici
nommé~\texttt{v}, est alors ajouté aux arguments lors de la traduction en
Viper.

Bien que l'implémentation en OCaml soit simple, il est nécessaire de
détailler les successions de pliages et de dépliages des prédicats de
représentation, ainsi que l'appel à \texttt{CellSegTrans}.
Nous détaillons
dans le code suivant le cas où une valeur est ajoutée à une
file vide.

\begin{minipage}[t]{0.5\textwidth}
  \centering
  \begin{gospel}
  let cell = Cons {
    content = x;
    next = Nil
  } in
  match q.last with
  | Nil ->
    q.length <- 1;
    q.first  <- cell;
    q.last   <- cell;
    (*@ fold cellSeg Nil empty Nil *)
    (*@ fold cellSeg cell
             (singleton x) Nil *)
    (*@ fold cellSeg q.first empty
             q.last *)
\end{gospel}
\end{minipage}
\hfill
\begin{minipage}[t]{0.5\textwidth}
  \centering
  \begin{viperlang}
  var c : Ref := new(content, next)
  c.content := x
  c.next := Nil()
  var cell : Cell := Cons(c)
  if (q.last.isNil) {
    q.length := 1
    q.first  := cell
    q.last   := cell
    fold CellSeg(Nil(), Seq[Int](),
                 Nil())
    fold CellSeg(q.last, Seq(x), Nil())
    fold CellSeg(q.first, Seq[Int](),
                 q.last)
  }
  \end{viperlang}
\end{minipage}
Notons que le filtrage de la valeur \texttt{Nil} est traduit vers le
destructeur \texttt{isNil} en Viper. La séquence de trois opérations
\texttt{fold} nous permet de montrer la validité du prédicat
\texttt{queue} quand la file possède un seul élément. Les deux
premiers \texttt{folds} sont utilisés pour montrer qu'entre
\texttt{q.last} et \texttt{Nil} il existe effectivement un segment de
liste contenant seulement l'élément \texttt{x}. Le dernier appel à
\texttt{fold} montre qu'entre \texttt{q.first} et \texttt{q.last} il y
a un segment de liste vide.

Pour ce qui est du cas de la file non vide, la preuve s'achève par un
appel au lemme \texttt{CellSeg\_trans} comme suit :

\vspace{-.5em}
\begin{minipage}[t]{0.5\textwidth}
  \centering
  \begin{gospel}
  | Cons last -> ...
    (* cascade de [fold]s *)
    (*@ apply CellSeg_trans q.first
          q.last cell (drop_last v)
          (take_last v) (singleton x) *)
    q.last <- cell
  \end{gospel}
\end{minipage}
\hfill
\begin{minipage}[t]{0.5\textwidth}
  \centering
  \begin{gospel}
  else { ...
    // cascade de [fold]s
    CellSeg_trans(q.first, q.last,
                  cell, drop_last(v),
    q.last <- cell
  }
  \end{gospel}
\end{minipage}
La fonction logique \texttt{drop\_last} renvoie une nouvelle séquence
sans le dernier élément de la séquence passée en argument tandis que la
fonction \texttt{take\_last} renvoie une séquence qui ne contient que
le dernier élément de son argument.
Avec le lemme \texttt{CellSeg\_trans}, nous obtenons la
possession d'un segment de liste entre \texttt{q.first} et
\texttt{cell} avec la séquence d'éléments~\texttt{v}
(\texttt{drop\_last(v) ++ take\_last(v) = v}). Enfin, avec ce résultat
et la dernière affectation \texttt{q.last <- cell}, nous parvenons à
rétablir le prédicat de représentation de la file étendue avec un
nouvel élément. Nous omettons ici la \textquote{cascade} d'opérations
\texttt{fold} qui nous permet de reconstruire des segments de
liste entre \texttt{q.first}, \texttt{q.last} et \texttt{cell},
réunissant les permissions nécessaires pour l'appel au lemme auxiliaire.


\paragraph{Concaténation de deux files.} Nous finissons notre étude de
cas par la concaténation de deux files. Dans le module \texttt{Queue},
la fonction qui concatène deux files est nommée
\texttt{transfer}. L'appel \texttt{transfer q1 q2} va concaténer les
éléments de~\texttt{q1} à la fin de~\texttt{q2}, en vidant par la même
occasion le contenu de \texttt{q1}. En pratique, il s'agit tout
simplement de relier le champ \texttt{last} de~\texttt{q2} au champ
\texttt{first} de \texttt{q1} et puis d'affecter \texttt{q2.last} à
\texttt{q1.last}. Cette opération illustre tout à fait l'intérêt de
maintenir des pointeurs vers le premier et dernier élément de la
file puisque nous arrivons à concaténer deux files en temps constant.
La spécification de la fonction \texttt{transfer} est tout aussi
directe, comme illustré ci-dessous :

\begin{minipage}[t]{.5\textwidth}
  \centering
  \begin{gospel}
let transfer
  (q1: queue) (q2: queue) =...
(*@ transfer q1 q2 [v1: int sequence]
             [v2: int sequence]
    requires queue q1 v1 && queue q2 v2
    ensures  queue q1 empty
    ensures  queue v2 (v2 ++ v1) *)
  \end{gospel}
\end{minipage}
\hfill
\begin{minipage}[t]{.5\textwidth}
  \centering
  \begin{viperlang}
method transfer (q1: Ref, q2: Ref,
  v1: Seq[Int], v2: Seq[Int])
  requires Queue(q1, v1) && Queue(q2, v2)
  ensures  Queue(q1, Seq[Int]())
  ensures  Queue(q2, v2 ++ v1)
{ ... }
  \end{viperlang}
\end{minipage}
En pré-condition, nous exigeons la possession des files~\texttt{q1}
et~\texttt{q2} avec respectivement les modèles~\texttt{v1} et~\texttt{v2}.
À la fin de l'exécution, nous conservons la
possession de~\texttt{q1} et~\texttt{q2}
mais leurs modèles respectifs sont mis à jour.
Nous nous concentrons sur le cas le plus intéressant de cette fonction :
quand~\texttt{q1} et~\texttt{q2} sont toutes deux non vides. Le
code OCaml, sa spécification Gospel, ainsi que la traduction obtenue en
Viper sont donnés ci-après.

\begin{minipage}[t]{.5\textwidth}
  \centering
  \begin{gospel}
  if q1.length > 0 then
    match q2.last with
    | Nil -> ...
    | Cons last ->
      q2.length <- q2.length + q1.length;
      (*@ unfold cellSeg q1.last
                 (take_last v1)
                 Nil *)
      last.next <- q1.first;
      (* cascade de [fold]s *)
      (*@ fold ... *)
      (*@ apply CellSeg_trans q1.last
                q.first q.last
                (take_last v1)
                (drop_last v)
                (take_last v) *)
      (*@ apply CellSeg_trans
                q1.first q1.last
                q.last (drop_last v1)
                ((take_last v1)
                  ++ (drop_last v))
                (take_last v) *)
      q2.last <- q1.last;
      clear q1
  \end{gospel}
\end{minipage}
\hfill
\begin{minipage}[t]{.5\textwidth}
  \centering
  \begin{viperlang}
  if (q1.length > 0) {
    if (q2.last.isNil) { ... }
    else {
      q2.length := q2.length + q1.length
      unfold CellSeg(q2.last,
                     take_last(v1),
                     Nil())
      q2.last.cell.next := q1.first
      // cascade de [fold]s
      fold ...
      CellSeg_trans(q2.last, q1.first,
                    q1.last,
                    take_last(v1),
                    drop_last(v),
                    take_last(v))
      CellSeg_trans(q2.first,
                    q2.last, q1.last,
                    drop_last(v1),
                    take_last(v1) ++
                      drop_last(v),
                    take_last(v))
      q2.last := q1.last
      clear(q1)
    }
  \end{viperlang}
\end{minipage}
Commençons par noter que le filtrage de la valeur \texttt{Cons last}
correspond en Viper, à la branche \texttt{else}. Dans ce cas, nous
sommes sûrs que la condition \texttt{q2.last.isCons} se vérifie.
La preuve s'appuie sur deux appels successifs au lemme
\texttt{CellSeg\_trans}.
Le premier permet d'établir l'existence d'un segment de liste
entre \texttt{q2.last} et \texttt{q1.last}.
Le second utilise le segment construit par l'appel précédent et
rend la possession d'un segment entre \texttt{q2.first} et
\texttt{q1.last}, ce qui permet d'établir la validité de file
concaténée. Finalement, l'appel à \texttt{clear} a pour but de vider
le contenu de la file~\texttt{q1}, respectant ainsi la post-condition
de la fonction \texttt{transfer}.

\section{Conclusion et perspectives}

Dans cet article, nous avons présenté une extension de l'outil
Cameleer permettant de traduire des programmes OCaml annotés avec
Gospel vers le langage intermédiaire Viper. Cette traduction ajoute un
\textit{backend} de logique de séparation à Cameleer afin de permettre
la vérification formelle de programmes manipulant des structures
allouées sur le tas. Nous avons présenté les nouveaux éléments ajoutés
au langage Gospel et pour chacun, détaillé comment notre
extension les transmet lors de la traduction vers Viper. Nous avons
illustré l'application de cette extension par la preuve formelle de
l'implémentation de certaines fonctions du module \texttt{Queue}
d'OCaml.
Il serait pertinent de tester la traduction sur d'autres structures de
données d'OCaml, comme les tables de hachage. Cela permettrait de
varier les types utilisés et d'identifier les limites potentielles de
l'extension actuelle.
L'ajout d'autres éléments de spécification (\emph{e.g.},
multiensemble), le polymorphisme d'OCaml et d'autres structures de
contrôle afin d'étendre les possibilités de programmes prouvables par
ce \emph{backend} est aussi envisagé.
Finalement, il serait intéressant d'étudier la
génération automatique de certaines opérations \texttt{fold} et
\texttt{unfold}. Actuellement, ces opérations doivent être insérées
manuellement dans le code OCaml, ce qui peut être à la fois laborieux et
source d'erreurs
. L'automatisation de cette
étape permettrait de simplifier la vérification, et des pistes dans
cette direction sont d'ores et déjà l'étude.

\bibliographystyle{alpha-fr}
\bibliography{gospel2viper}

\end{document}